\documentstyle[preprint,aps]{revtex}

\newcommand{\beq}{\begin{equation}}
\newcommand{\enq}{\end{equation}}

\newcommand{\Si}[1]{\mbox{Si$_{#1}$}}

\begin{document}
\draft
\title{Magic Numbers of Silicon Clusters}
\author{Jun Pan\footnote{The Department of Physics,
New York University, New York, NY 10003-6621.} and Mushti V. Ramakrishna}
\address{The Department of Chemistry, New York University,
New York, NY 10003-6621.}
\date{Submitted to Phys. Rev. Lett., \today}
\maketitle

\begin{abstract}

A structural model for intermediate sized silicon clusters is proposed
that is able to generate unique structures without any dangling bonds.
This structural model consists of bulk-like core of five atoms
surrounded by fullerene-like surface.  Reconstruction of the ideal
fullerene geometry results in the formation of crown atoms surrounded
by $\pi$-bonded dimer pairs.  This model yields unique structures for
\Si{33}, \Si{39}, and \Si{45} clusters without any dangling bonds and
hence explains why these clusters are least reactive towards
chemisorption of ammonia, methanol, ethylene, and water.  This model is
also consistent with the experimental finding that silicon clusters
undergo a transition from prolate to spherical shapes at \Si{27}.
Finally, reagent specific chemisorption reactivities observed
experimentally is explained based on the electronic structures of the
reagents.

\end{abstract}
\pacs{PACS numbers: 36.40+d, 61.43.Bn, 61.46.+w, 68.35.Bs, 82.65.My}

Smalley and co-workers carried out extensive studies on chemisorption
of various reagents on intermediate sized silicon clusters
\cite{Elkind:87}.   These studies revealed that the reactivity rates
for ammonia (NH$_3$) chemisorption vary by over three orders of
magnitude as a function of cluster size, with 21, 25, 33, 39, and 45
atom clusters being particularly unreactive.  Clusters containing more
than forty seven atoms are highly reactive and they do not display
strong oscillations in reactivities as a function of the cluster size.
Similar results were obtained with methanol (CH$_3$OH), ethylene
(C$_2$H$_4$), and water (H$_2$O).  On the other hand, nitric oxide (NO)
and oxygen (O$_2$) were found to react readily with all clusters in
this size range.

Several structural models have been proposed to explain these
experimental data
\cite{Phillips:88,Jelski:88,Kaxiras:89,Patterson:90,Swift:91}.
However, these models do not yield unique and consistent structures for
different cluster sizes and each cluster structure has to be obtained
independently.  Furthermore, these models do not satisfy the essential
criterion that the structures of the unreactive clusters should not
have any dangling bonds.

In this Letter we propose a consistent model that generates the
structures of the intermediate sized unreactive silicon clusters in a
systematic way.  The structures thus generated are unique for \Si{33},
\Si{39}, and \Si{45} clusters.  Furthermore, these clusters do not have
any dangling bonds and hence explains why these clusters are unreactive
for chemisorption.  Finally, our model is consistent with the
experimental finding of Jarrold and Constant that silicon clusters
undergo a shape transition from prolate to spherical shapes at \Si{27}
\cite{Jarrold:91}.

Our structural model for silicon clusters consists of bulk-like core of
five atoms surrounded by fullerene-like surface.  The core atoms bind
to the twelve surface atoms, thus rendering them bulk-like with
four-fold coordination.  The surface then relaxes from its ideal
fullerene geometry to give rise to crown atoms and dimer pairs.  The
crown atoms are called adatoms in surface science literature
\cite{crown}.  These crown atoms are formally three-fold coordinated
and possess one dangling bond each.  The dimer pairs are also formally
three-fold coordinated, but they eliminate their dangling bonds through
local $\pi$ bonding.  This model yields structures with seventeen
four-fold coordinated atoms, four crown atoms, and the rest dimer
pairs.  The essential feature of this construction
is that the bulk-like core and fullerene-like surface make these
structures stable.  This model is applicable to clusters containing
more than twenty atoms.

Unlike carbon, silicon does not form strong delocalized $\pi$ bonds.
Consequently, fullerene cage structures \cite{Curl:91,Boo:92}, which
require strong delocalized $\pi$ bonds, are energetically unfavorable
for silicon.  For this reason, silicon clusters prefer as few surface
atoms as possible.  Nonetheless, the fullerene geometry for the
surface, consisting of interlocking pentagons and hexagons, gives
special stability to the surface atoms.  Furthermore, since delocalized
$\pi$ bonding is not favorable in silicon, we expect the surface atoms
to relax from their ideal fullerene geometry to allow for dimer
formation through strong local $\pi$ bonding.  Our model accounts for
all these facts.

The 5-atom core in our model has the exact structure of bulk silicon
with one atom in the center bonded to four atoms arranged in a perfect
tetrahedral symmetry.  There are numerous ways to orient the 5-atom
core inside the fullerene cage and bind it to the surface atoms.
Furthermore, structural isomers may exist for fullerenes of any size
\cite{Boo:92,Fowler:92}.  Thus it is possible to use this model to
generate all possible structural isomers for any odd numbered
intermediate sized cluster.  However, a particular orientation of the
5-atom core inside the fullerene cage will yield structures with
maximum number of dimer pairs and least number of dangling bonds.  Such
isomers are likely to be most stable.

The 28- and 40-atom fullerenes are the only ones belonging to the
symmetry group $T_d$ in the 20-60 atom size range \cite{Boo:92}.  We
generate the \Si{33} and \Si{45} structures by inserting the 5-atom
core inside the 28- and 40-atom fullerene, respectively.  We orient the
5-atom pyramid in such a way that the central atom (A, violet), the
apex atom (B, blue) of the pyramid, and the crown atom (C, red) on the
surface lie on a line.  The C atom is the central atom shared by three
fused pentagons.  This atom is surrounded by three other surface atoms
(D, green), which relax inwards to bind to the four core atoms B.  This
relaxation motion is identical to that necessary to form the 2 $\times$
1 reconstruction on the bulk Si(111) surface
\cite{Cohen:84,Lannoo:91}.  With an activation barrier of only
$\approx$ 0.01 eV \cite{Cohen:84,Northrup:82} and gain of 2.3 eV/bond
due to $\sigma$ bond formation between B and D
\cite{Lannoo:91,Brenner:91}, such a relaxation of fullerene surface is
feasible even at 100 K.  Finally, the remaining surface atoms (E,
orange) readjust to form as many dimers as possible.  This relaxation
is similar to that on Si(100) surface yielding dimer pairs
\cite{Lannoo:91,Chadi:79}.  The dimers are $\sigma$-bonded pair of
atoms whose dangling bonds are saturated through strong local $\pi$
bonds.  Because of the tetrahedral symmetry of the core as well as of
the reconstructed fullerene cage, the final \Si{33} and \Si{45}
structures also have tetrahedral symmetry.
The \Si{39} structure is also generated in this way, starting from the
34-atom fullerene cage and stuffing five atoms inside it.  However, the
structure thus generated has only C$_{3v}$ symmetry, because of the
lower symmetry of the fullerene cage \cite{Boo:92}.

The \Si{33}, \Si{39}, and \Si{45} cluster structures thus generated are
displayed in Fig. 1.  These structures have seventeen four-fold
coordinated atoms, four crown atoms, and variable number of dimer
pairs.  The \Si{33} structure has six dimer pairs, whereas \Si{45} has
twelve dimer pairs forming four hexagons.  On the other hand, the
surface of the \Si{39} cluster consists of nine dimer pairs, three of
which form a hexagon.

We constructed \Si{35} and \Si{43} clusters also using the proposed
model, but found that these structures do not possess the
characteristic crown-atom-dimer pattern found in unreactive clusters.
Consequently, these structures possessed large number of dangling
bonds.  This explains the highly reactive nature of these clusters.

In principle, the proposed model can be used to construct \Si{25}
structure also.  However, \Si{20} cage is too small to accommodate five
atoms inside it without undue strain.  Consequently, \Si{25} will
prefer a different geometry.  Indeed, there is experimental evidence
that clusters smaller than \Si{27} do not possess spherical shapes
characteristic of larger clusters; instead they seem to prefer prolate
shapes \cite{Jarrold:91}.  Our inability to generate a compact
structure for \Si{25} explains this experimental observation also.

We verified our model by constructing structures of all the clusters
discussed here using the molecular modelling kits \cite{Jones:94}.
These structures unambiguously demonstrated that only 33-, 39-, and
45-atom clusters possess the crown-atom-dimer pattern on their
surfaces.  We then calculated the atomic coordinates of these clusters
from the tetrahedral geometry of the 5-atom core and the known
geometries of the fullerene structures \cite{Boo:92}, assuming that all
bond lengths are equal to the bulk value of 2.35 \AA.   The structures
thus generated are displayed in Fig. 1.  We then relaxed these
structures by carrying out molecular dynamics at 100 K using the
Kaxiras-Pandey potential \cite{Kaxiras:88}.   The final structures
obtained from these simulations are nearly identical to the initial
ones.  This indicates that the proposed structures are locally stable.

The computer generated structures displayed in Fig. 1 revealed that the
crown atoms are able to form a fourth bond to the core atoms B, thus
rendering the B atoms formally five-fold coordinated.  The B-C bond
arises from the back donation of the electrons from C to B and it
weakens the neighboring bonds through electronic repulsion.  The
dangling bond on the crown atom is thus eliminated and these magic
number clusters become unreactive.

The classical potentials available at present
\cite{Kaxiras:88,Stillinger:85} are most appropriate for bulk silicon
and related structures.  Such potentials may not be suitable for
describing unusual bonding patterns, such as the five-fold coordinated
silicon atoms found in these clusters.  Consequently, we cannot use the
available classical potentials to prove that the proposed structures
correspond to the lowest energy structures.  Finally, structural
determination based on the first principles electronic structure
calculations are extremely demanding computationally at present for
such large clusters, especially if adequate basis sets and grid sizes
are employed and the calculations are converged to better than 0.01 eV
accuracy.  Furthermore, such complicated calculations do not provide
the conceptual framework for understanding cluster properties.  On the
other hand, our simple physically motivated stuffed fullerene model
yields insights into the nature of bonding in silicon clusters and
explains the experimental trends in reactivities.

There exist several competing structural models
\cite{Phillips:88,Jelski:88,Kaxiras:89,Patterson:90,Swift:91}
of silicon clusters that attempt to explain the experimental reactivity
data of Smalley and co-workers \cite{Elkind:87}.
For example, Kaxiras has proposed structures of Si$_{33}$ and Si$_{45}$
clusters based on the reconstructed 7 $\times$ 7 and 2 $\times$ 1
surfaces of bulk Si(111), respectively \cite{Kaxiras:89}.
However, this model does not explain the reactivity data since bulk
surfaces are highly reactive.  Furthermore, it is inconsistent that the
surface of \Si{45} should be the metastable 2 $\times$ 1 surface rather
than the more stable 7 $\times$ 7 surface \cite{Lannoo:91}.  Finally,
the \Si{45} structure of Kaxiras has forty dangling bonds, which make
it highly reactive, contrary to the experiments.  To overcome this
discrepancy between experiment and theory, Kaxiras has proposed that
the dangling bonds on \Si{45} form $\pi$-bonded chains, analogous to
the 2 $\times$ 1 reconstruction of the bulk Si(111) surface
\cite{Pandey:81}.  However, silicon favors strong local $\pi$ bonds
over delocalized $\pi$-bonded chains.  This is the reason why the 2
$\times$ 1 reconstruction, involving $\pi$-bonded chains
\cite{Pandey:81}, is metastable with respect to the locally
$\pi$-bonded 7 $\times$ 7 reconstruction on the bulk Si(111) surface
\cite{Lannoo:91}.  For the same reason, the \Si{45} structure proposed
by Kaxiras is a metastable and highly reactive structure.  Indeed,
Jelski and co-workers disputed the Kaxiras model of \Si{45} by
constructing alternative structures for Si$_{45}$ that are lower in
energy, but do not possess any of the features of the reconstructed
bulk surfaces \cite{Swift:91}.

The bonding characteristics of silicon differ in subtle ways from that
of carbon.  In carbon, delocalized $\pi$ bonds are favored over local
$\pi$ bonds, whereas the opposite is true in silicon.  For this reason,
graphite is the most stable form of carbon at room temperature and
atmospheric pressure, but not the graphite form of silicon.  Likewise,
the bulk (111) surface of the diamond form of carbon exhibits the 2
$\times$ 1 reconstruction, but not the 7 $\times$ 7 reconstruction
\cite{Pate:86,Bokor:86}.  These examples, illustrate how subtle
differences in bonding characteristics determine possible crystal
structures and surface reconstructions.  The same is true of clusters
and the models of cluster structures should account for these
characteristics.  Our model of silicon clusters accounts for
these facts by focussing on structures that are able to form maximum
number of $\sigma$ bonds and eliminate their surface dangling bonds
through local $\pi$ bonding.

Our structure for \Si{33} is identical to that proposed by Kaxiras
\cite{Kaxiras:89} and Patterson and Messmer \cite{Patterson:90}.  This
structure has been shown to be locally stable \cite{Feldman:91}.  But
our \Si{45} structure is different from that of Kaxiras
\cite{Kaxiras:89}.  However, we can generate the \Si{45}
structure of Kaxiras by stuffing one atom inside a 44-atom fullerene
cage and allowing for the reconstruction of the fullerene surface.
Thus our model is very general, subsuming the Kaxiras model within it.

The reactivity patterns of NO and O$_2$ are different from those found
for NH$_3$, CH$_3$OH, C$_2$H$_4$, and H$_2$O \cite{Elkind:87}.  This
may be explained based on the ground state electronic structures of
these reagents.  NH$_3$, CH$_3$OH, C$_2$H$_4$, and H$_2$O in their
ground states have closed shell electronic structure with all electrons
paired.  On the other hand, NO and O$_2$ in their ground states are
$^2\Pi_g$ and $^3\Sigma_g^-$, possessing one and two unpaired
electrons, respectively \cite{Herzberg:50}.  Consequently, NH$_3$,
CH$_3$OH, C$_2$H$_4$, and H$_2$O can chemisorb only at those sites
where excess electron density is present due to dangling bonds.  Such a
selectivity gives rise to highly oscillatory pattern in the
reactivities, because the number of dangling bonds varies as a function
of cluster size.  The magic number clusters are unreactive because
they do not possess any dangling bonds.  On the other hand, NO
$(^2\Pi_g)$ and O$_2$ $(^3\Sigma_g^-)$ can chemisorb anywhere, because
these reagents carry the necessary dangling bonds for instigating the
reaction anywhere on the cluster surface.  Hence, NO and O$_2$ readily
react with all clusters and do not display the oscillatory pattern in
their chemical reactivities.  This explains the reagent specific
chemisorption reactivities observed experimentally \cite{Elkind:87}.

The magic number clusters are not completely inert towards the closed
shell reagents \cite{Elkind:87}.  These clusters are more reactive
towards  NH$_3$, CH$_3$OH, and H$_2$O than towards C$_2$H$_4$.  This is
because NH$_3$, CH$_3$OH, and H$_2$O have lone pairs on either nitrogen
or oxygen and these lone pairs have a small probability of instigating
reaction on the cluster surface.    A lone pair is a pair of electrons
that is not part of a bond.  C$_2$H$_4$ does not have any lone pairs
and hence the magic number clusters are quite unreactive towards this
molecule.  The electronic the structure of reagents thus explains even
subtle variations in the reactivities of magic number clusters towards
a group of related reagents.

In summary, we propose a structural model for the unreactive silicon
clusters containing more than twenty atoms.   This model consists of
bulk-like core of five atoms surrounded by reconstructed fullerene
surface.  The resulting structures for \Si{33}, \Si{39}, and \Si{45}
are unique, have maximum number of four-fold coordinated atoms, minimum
number of surface atoms, and zero dangling bonds.  Such unique
structures cannot be built for other intermediate sized clusters and
hence they will have larger number of dangling bonds.  This explains
why \Si{33}, \Si{39}, and \Si{45} clusters are least reactive towards
closed shell reagents ammonia, methanol, ethylene, and water
\cite{Elkind:87}.  Our model also indicates that \Si{25} cluster cannot
be formed in a spherical shape.  This result is consistent with the
experimental finding that silicon clusters undergo a shape transition
from prolate to spherical shapes at \Si{27} \cite{Jarrold:91}.
Finally, two distinct patterns of chemisorption reactivities observed
experimentally are explained based on the electronic structures of the
reagents.  The reactivities of closed shell reagents depend on the
available number of dangling bond sites, whereas the reactivities of
free radical reagents are not so dependent.  Consequently, only the
closed shell reagents are sensitive to the cluster structure and hence
exhibit the highly oscillatory pattern in reactivities as a function of
the cluster size.

This research is supported by the New York University and the Donors of
The Petroleum Research Fund (ACS-PRF \# 26488-G), administered by the
American Chemical Society.

\begin{figure}
\caption{Structures of \Si{33}, \Si{39}, and \Si{45} clusters obtained
using the proposed stuffed fullerene model.   The atoms in different
chemical environments are colored differently from violet to red, while
only the representative atoms are labelled from A to E.  These clusters
do not possess any dangling bonds and hence are least reactive towards
the closed shell reagents ammonia, methanol, ethylene, and water. }

\end{figure}

\begin{thebibliography}{1}

\bibitem[1]{Elkind:87} J. L. Elkind, J. M. Alford, F. D. Weiss,
R. T. Laaksonen, and R. E. Smalley, J. Chem. Phys. {\bf 87}, 2397 (1987);
S. Maruyama, L. R. Anderson, and R. E. Smalley,
J. Chem. Phys. {\bf 93}, 5349 (1990);
J. M. Alford, R. T. Laaksonen, and R. E.
Smalley, J. Chem. Phys. {\bf 94}, 2618 (1991);
L. R. Anderson, S. Maruyama, and R. E. Smalley,
Chem. Phys. Lett. {\bf 176}, 348 (1991).

\bibitem[2]{Phillips:88} J. C. Phillips, J. Chem. Phys. {\bf 88}, 2090
(1988).

\bibitem[3]{Jelski:88} D. A. Jelski, Z. C. Wu, and T. F. George,
Chem. Phys. Lett. {\bf 150}, 447 (1988).

\bibitem[4]{Kaxiras:89} E. Kaxiras, Chem. Phys. Lett. {\bf 163}, 323,
(1989); Phys. Rev. Lett. {\bf 64}, 551 (1990).

\bibitem[5]{Patterson:90} C. H. Patterson and R. P. Messmer,
Phys. Rev.  B {\bf 42}, 7530 (1990).

\bibitem[6]{Swift:91} B. L. Swift, D. A. Jelski, D. S. Higgs, T. T. Rantala,
and T. F. George, Phys. Rev. Lett. {\bf 66}, 2686 (1991);
D. A. Jelski, B. L. Swift, T. T. Rantala, X. Xia, T. F. George,
J. Chem. Phys. {\bf 95}, 8552 (1991).

\bibitem[7]{Jarrold:91} M. F. Jarrold and V. A. Constant,
Phys. Rev. Lett. {\bf 67}, 2994 (1991).

\bibitem[8]{crown} The word adatom connotes adsorbed atom, thereby
making one think that this is a weakly bonded atom.  In reality, this
atom is strongly bonded to the cluster.  Consequently, we prefer to
call it the crown atom, since it is at the center of three fused
polygons.

\bibitem[9]{Curl:91} R. F. Curl and R. E. Smalley,
Sci. Am. {\bf 10}, 54, (1991).

\bibitem[10]{Boo:92} W. O. J. Boo, J. Chem. Education {\bf 69}, 605 (1992).

\bibitem[11]{Fowler:92} P. W. Fowler, in {\em FULLERENES: Status and
Perspectives}, eds. C. Taliani, G. Ruani, and R. Zamboni,
(World Scientific, Singapore, 1992).

\bibitem[12]{Cohen:84} M. L. Cohen and S. G. Louie,
Ann. Rev. Phys. Chem. {\bf 35}, 537 (1984).

\bibitem[13]{Lannoo:91} M. Lannoo and P. Friedel,
{\em Atomic and Electronic Structure of Surfaces},
Ch. 4, (Springer-Verlag, Berlin, 1991).

\bibitem[14]{Northrup:82} J. E. Northrup and M. L. Cohen,
Phys. Rev. Lett. {\bf 49}, 1349 (1982).

\bibitem[15]{Brenner:91} D. W. Brenner, B. I. Dunlap, J. A. Harrison,
J. W. Mintmire, R. C. Mowrey, D. H. Robertson, and C. T. White,
Phys. Rev. B {\bf 44}, 3479 (1991).

\bibitem[16]{Chadi:79} D. J. Chadi,
Phys. Rev. Lett. {\bf 43}, 43 (1979).

\bibitem[17]{Jones:94} {\em Foundation Set for General and Organic
Chemistry}, (Jones and Bartlett Publishers, Boston, 1994).

\bibitem[18]{Kaxiras:88} E. Kaxiras and K. C. Pandey,
Phys. Rev. B 38, 12736 (1988).

\bibitem[19]{Stillinger:85}
F. Stillinger and T. Weber, Phys. Rev. B {\bf 31}, 5262 (1985).
R. Biswas and D. R. Hamann, Phys. Rev. Lett. {\bf 55}, 2001 (1985);
Phys. Rev. B {\bf 36}, 6434 (1987);
J. Tersoff, Phys. Rev. Lett. {\bf 56}, 632 (1986);
P. C. Kelires and J. Tersoff, Phys. Rev. Lett. {\bf 61}, 562 (1988);
B. C. Bolding and H. C. Andersen, Phys. Rev. B {\bf 41}, 10568 (1990);
J. R. Chelikowsky, K. M. Glassford, and J. C. Phillips,
Phys. Rev. B {\bf 44}, 1538 (1991).

\bibitem[20]{Pandey:81} K. C. Pandey,
Phys. Rev. Lett. 47, 1913 (1981);
Phys. Rev. B {\bf 25}, 4338 (1982).

\bibitem[21]{Pate:86} B. B. Pate, Surf. Sci. {\bf 83}, 165 (1986).

\bibitem[22]{Bokor:86} J. Bokor, R. Storz, R. R. Freeman, and P. H. Bucksbaum,
Phys. Rev. Lett. {\bf 57}, 881 (1986).

\bibitem[23]{Feldman:91} J. L. Feldman, E. Kaxiras, and X.-P. Li,
Phys. Rev. B {\bf 44}, 8334 (1991).

\bibitem[24]{Herzberg:50} G. Herzberg, {\em Molecular Spectra and Molecular
Structure I. Spectra of Diatomic Molecules},
(Van Nostrand, New York, 1950), page 343.

\end{thebibliography}
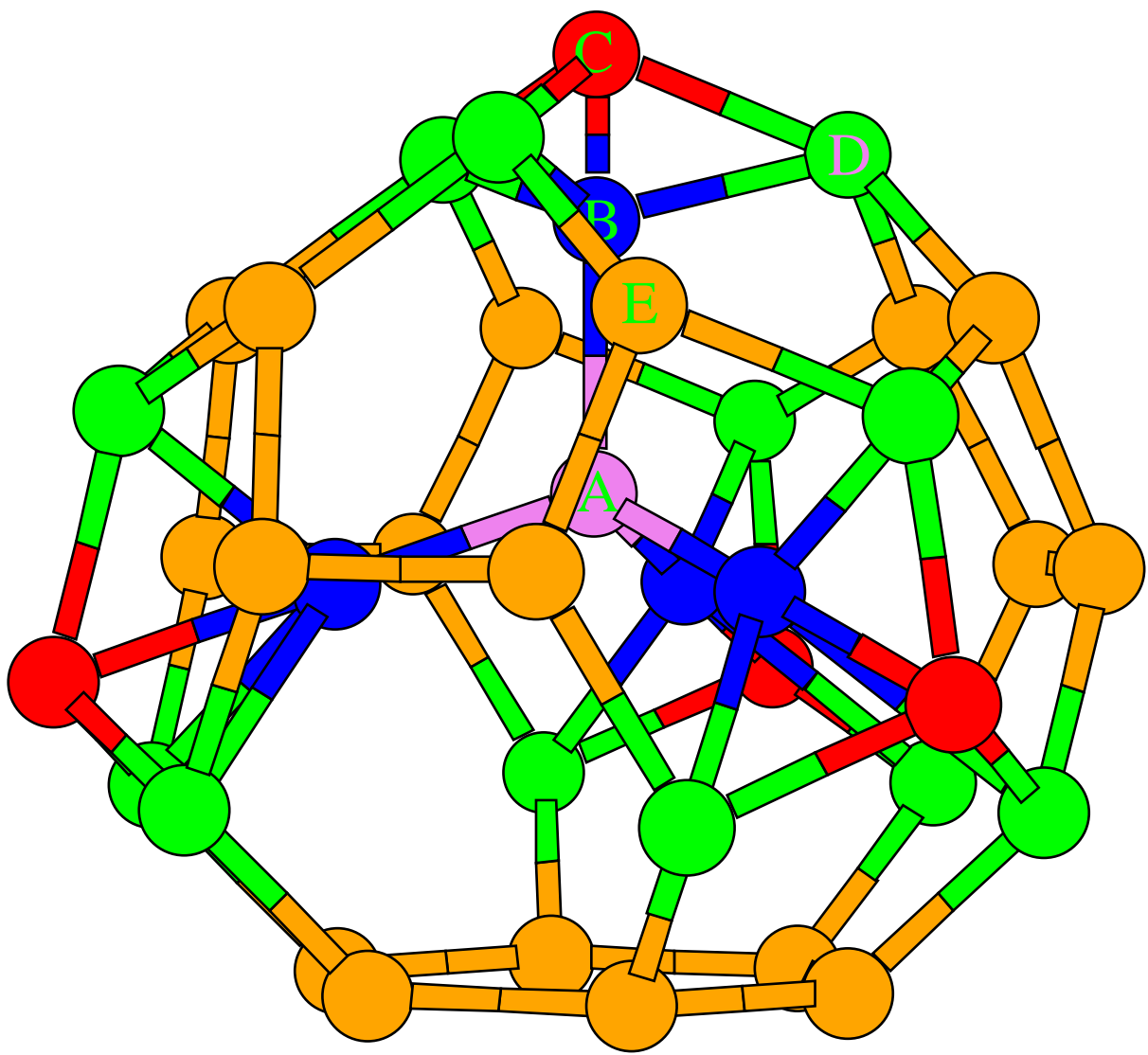
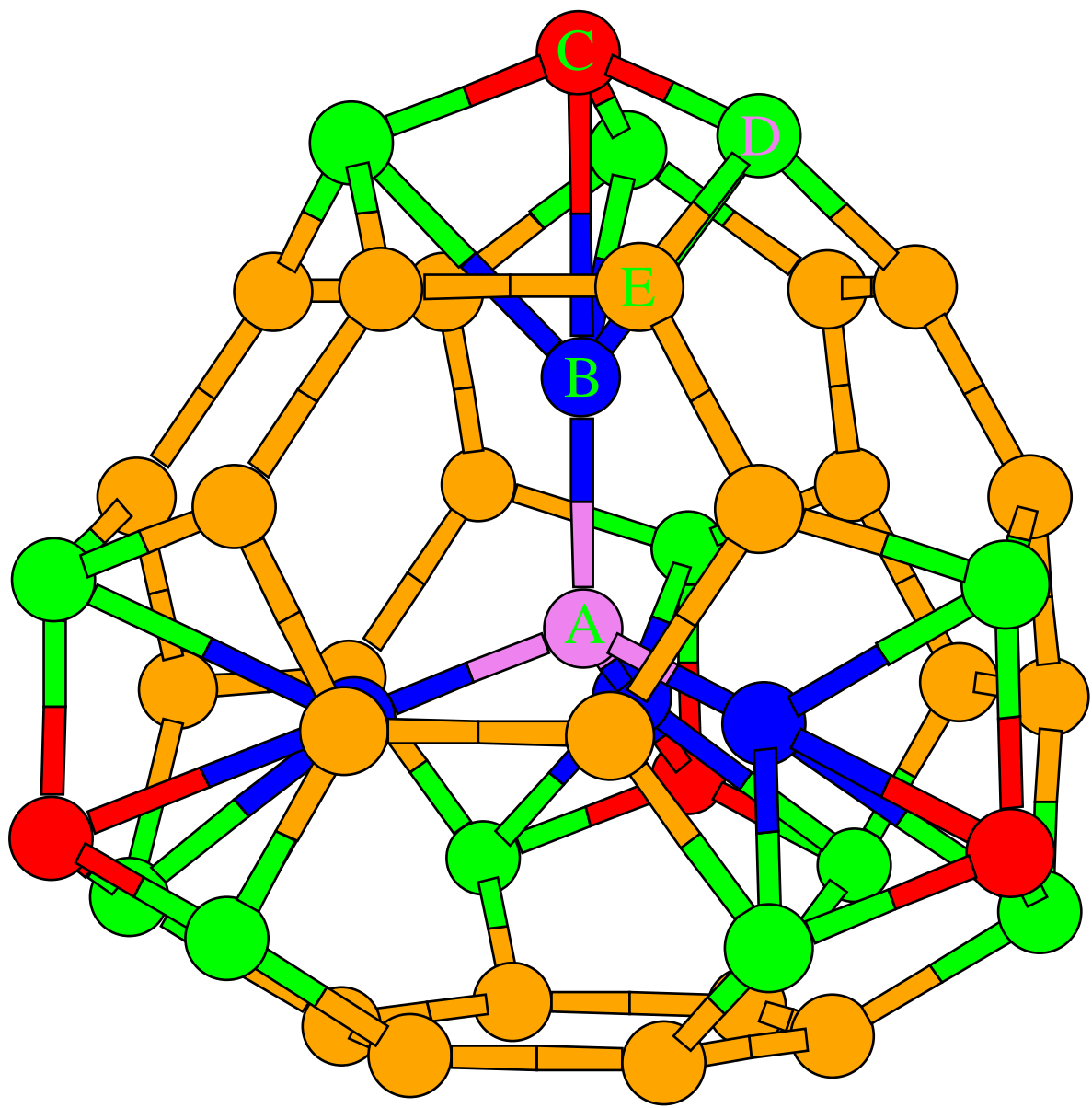
\end{document}